\documentclass[aps,floatfix,superscriptaddress,twocolumn,tightenlines,amsmathm,textcomp,amssymb,nofootinbib,10pt,nofootinbib]{revtex4-1}
\usepackage[utf8]{inputenc}
\usepackage[T1]{fontenc}
\usepackage{braket}
\usepackage{graphicx}
\usepackage{booktabs}
\usepackage{mathtools}	
\usepackage{amssymb}
\usepackage{amsmath}	
\usepackage{amsfonts}
\usepackage{mathrsfs}
\usepackage{color}
\usepackage{upgreek}
\usepackage{amsthm}
\usepackage{psfrag}
\usepackage{ifsym}
\usepackage{dsfont}
\usepackage{enumerate}
\usepackage{enumitem}
\usepackage{multirow} 
\usepackage{lipsum}
\usepackage{float}
\usepackage[separate-uncertainty=true]{siunitx}
\usepackage{setspace}
\usepackage[dvipsnames]{xcolor}
\usepackage[normalem]{ulem}
\usepackage{textgreek}
\usepackage{tabularx}
\usepackage{silence}
\usepackage{multirow}
\usepackage{booktabs}

\usepackage{hyperref}
\definecolor{refColour}{RGB}{0, 0, 0}
\definecolor{citeColour}{RGB}{0, 0, 0}
\definecolor{extraColour}{RGB}{0, 0, 0}
\hypersetup{colorlinks=true,linkcolor=Magenta,citecolor=Magenta}

\usepackage{titlesec}
\titleformat*{\subsection}{\normalfont\centering}
\titleformat*{\section}{\normalfont\centering\large}

\newcommand{\one}{\leavevmode\hbox{\small1\normalsize\kern-.33em1}}

\usepackage{siunitx}
\usepackage{float}

\newcolumntype{Y}{>{\centering\arraybackslash}X}
\DeclareMathOperator*{\argmax}{\arg\,\max}

\makeatletter
\def\thickhline{%
  \noalign{\ifnum0=`}\fi\hrule \@height \thickarrayrulewidth \futurelet
   \reserved@a\@xthickhline}
\def\@xthickhline{\ifx\reserved@a\thickhline
               \vskip\doublerulesep
               \vskip-\thickarrayrulewidth
             \fi
      \ifnum0=`{\fi}}
\makeatother

\begin{document}
\title{\normalfont\Large{Optimised finite key for satellite up-link entangled quantum key distribution}}

\author{T. Jaeken*}
\affiliation{Institute of Photonics and Quantum Sciences, School of Engineering and Physical Sciences, Heriot-Watt University, Edinburgh EH14 4AS, UK}
\author{F. Redza*}
\email{fr24@hw.ac.uk}
\affiliation{Institute of Photonics and Quantum Sciences, School of Engineering and Physical Sciences, Heriot-Watt University, Edinburgh EH14 4AS, UK}

\author{S. B{\"a}uml}
\affiliation{Dahlem Center for Complex Quantum Systems, Freie Universit{\"a}t Berlin, 14195 Berlin, Germany}

\author{D. K. L. Oi}
\affiliation{SUPA Department of Physics, University of Strathclyde, Glasgow, G4 0NG, UK}

\author{A. Fedrizzi}
\affiliation{Institute of Photonics and Quantum Sciences, School of Engineering and Physical Sciences, Heriot-Watt University, Edinburgh EH14 4AS, UK}

\author{J. S. Sidhu*}
\email{jsmdrsidhu@gmail.com}
\affiliation{Dahlem Center for Complex Quantum Systems, Freie Universit{\"a}t Berlin, 14195 Berlin, Germany}
\affiliation{SUPA Department of Physics, University of Strathclyde, Glasgow, G4 0NG, UK}
\affiliation{School of Mathematical and Physical Sciences, University of Sheffield, Sheffield, S3 7RH, UK}

\def\thefootnote{*}\footnotetext{These authors contributed equally to this work}\def\thefootnote{\arabic{footnote}}

\begin{abstract}
Entanglement-based quantum key distribution from ground to satellite has been proposed as a scalable primitive for global quantum networks and will be demonstrated as part of the imminent QEYSSat mission. Both theoretical studies and experimental emulations have shown that a significant asymptotic key rate can be expected. In this work we investigate this scenario with a more rigorous security analysis taking into account the finite size effects that arise from short block lengths common in satellite transmission. Our findings suggest a slight alteration of the optimal source operation when informed by finite statistics as opposed to the asymptotic regime, notably that the pump power should be slightly lowered.
We quantify the annual key rate based on a dynamically optimised source operation for the Heriot-Watt ground station and several satellite parameters. We find an improved annual key rate when accumulating all raw key material before extracting a secure key instead of doing so for each overpass. We ultimately show that QEYSSat will be able to generate a significant secure annual key with both strategies.
\end{abstract}

\maketitle

\section{Introduction}
\label{sec:intro}

\noindent
Quantum networking is emerging as the foundational infrastructure for distributed quantum technologies~\cite{dowlingQuantumTechnologySecond2003}.
It enables physically separated quantum systems, including sensors~\cite{proctorMultiparameterEstimationNetworked2018,Sidhu2018_arxiv,sidhuTightBoundsSimultaneous2021,sidhuGeometricPerspectiveQuantum2020,brooksQuantumlimitedOpticalDelay2025,humphreysQuantumEnhancedMultiple2013,hoQuantumprivateDistributedSensing2024}, processors~\cite{knillSchemeEfficientQuantum2001}, and imaging devices to operate in a coherent and coordinated manner, and is unlocking transformative capabilities in secure communications~\cite{scaraniSecurityPracticalQuantum2009,pirandolaAdvancesQuantumCryptography2020}, distributed computing~\cite{fitzsimonsPrivateQuantumComputation2017a}, enhanced metrology, timing~\cite{komarQuantumNetworkClocks2014,ilo-okekeRemoteQuantumClock2018,jozsaQuantumClockSynchronization2000} and beyond. 
A characteristic capability of such a network is the ability to perform entanglement-based \emph{quantum key distribution} (QKD) protocols such as \emph{Bennett-Brassard-Mermin 1992} (BBM92) which we focus on here.
In particular, the achievable secure key rate provides a quantitative metric for link performance, directly informing the design and deployment of future quantum networks~\cite{Wallnofer2022,Sidhu2026operational}.

\emph{Satellite-based QKD} (SatQKD) plays a pivotal role in extending quantum communication links to global scales by exploiting free-space optical links through the space domain~\cite{yehiaConnectingQuantumCities2024,goswamiSatellitesPromiseGlobalscale2025a,deforgesdeparnySatellitebasedQuantumInformation2023,contiSatelliteTerrestrialQuantumNetworks2024}.
In recent years, significant experimental and theoretical advances have demonstrated the viability of SatQKD~\cite{sidhu2021key, Sidhu2022_npjQI, sidhu2023key, Sidhu2023finite}, with the Chinese Micius~\cite{luMiciusQuantumExperiments2022,chenIntegratedSpacetogroundQuantum2021a,renGroundtosatelliteQuantumTeleportation2017,liaoSatellitetogroundQuantumKey2017,yinSatellitetoGroundEntanglementBasedQuantum2017} and Jinan~\cite{liMicrosatellitebasedRealtimeQuantum2025} missions demonstrating entanglement distribution and quantum communication, respectively, over unprecedented distances and showcasing a range of fundamental quantum networking capabilities~\cite{Wang2026}.
SatQKD will serve as a foundational component for interconnecting ground-based quantum networks across continents, helping lay the architecture for a future quantum internet~\cite{wehnerQuantumInternetVision2018}.

Among SatQKD architectures, an up-link configuration, where quantum signals are transmitted from an \emph{optical ground station} (OGS) to a satellite, provides operational advantages. 
Up-link missions offer greater flexibility because the optical source and associated hardware remain on the ground, allowing upgrades, maintenance, and iteration without launching new payloads.
The satellite therefore hosts only the receiver, which has lower \emph{size, weight, and power} (SWaP) requirements than the source --- key metrics for satellite design and launch constraints. Up-link settings could also be used in measurement-device-independent \cite{lo2012measurement,braunstein2012side} settings where the satellite is untrusted.  
Despite these advantages, the up-link path presents distinct technical challenges, particularly due to the shower curtain effect~\cite{coneybear1978} of the high atmospheric turbulence encountered near the transmitter and the background noise seen by the satellite receiver that faces the Earth. 
These challenges severely constrain the achievable \emph{secure key length} (SKL), making performance optimisation critical. 
In this regime of small key sizes, security guarantees are highly dependent on considering the effect of finite statistics~\cite{Islam2024finite}.
Rigorous finite key analysis~\cite{Lim} is essential to accurately characterise performance and guide system-level optimisation.
Similar finite key studies have been performed into down-link entanglement-based QKD~\cite{wangScalabilityBottlenecksEntanglementbased2026,Sidhu2026operational}.

In this work, we confirm the feasibility of up-link satellite quantum communication with entangled photons by applying finite-key analysis~\cite{Lim} to the results obtained in our previous work~\cite{jaekenEmulationSatelliteUplink2025}. The finite-key analysis of~\cite{Lim} is based on the security proof of~\cite{tomamichel2017largely}, which uses entropic uncertainty relations to prove security against coherent attacks on the QKD system. We apply this framework to model system performance under realistic satellite dynamics and operational constraints, and we evaluate the achievable annual SKL from all overpasses above the minimum elevation angle of the OGS.
We then outline the hardware requirements and post-processing strategies needed in order to implement up-link quantum communication under these constraints.
This analysis addresses open questions surrounding the feasibility and optimisation of up-link SatQKD missions and supports ongoing efforts to mature the architecture of space-based quantum networks.
A succession of small-satellite quantum communication missions is expected to launch in the near term, specifically the QEYSSat~\cite{bourgoinComprehensiveDesignPerformance2013b,srikaraQuantumEntanglementDistribution2024,jenneweinQEYSSATMissionProposal2014b} mission which we use to guide our simulations.
Therefore, our results provide timely information on the potential capabilities and limitations of these missions.

%--

\section{Background}
\label{sec:background}

\noindent
To estimate the performance of an up-link SatQKD system, we require both a model of the optical channel and a framework for evaluating the resulting secure key length. In particular, the time-dependent channel loss determines the coincidence rate and error statistics over a satellite overpass. We therefore begin by modelling the elevation-dependent up-link loss in Sec.~\ref{sec:up-link}. Given these detection statistics, the achievable secure key length must then be evaluated within a finite-key security framework. This accounts for statistical fluctuations in parameter estimation and the requirements of composable security. We outline this analysis in Sec.~\ref{sec:secure_key_analysis}, where the secure key length is obtained from the observed coincidence rates and error probabilities.

%--

\subsection{Up-link channel}
\label{sec:up-link}

\begin{figure}
    \centering
    \includegraphics[width=\linewidth]{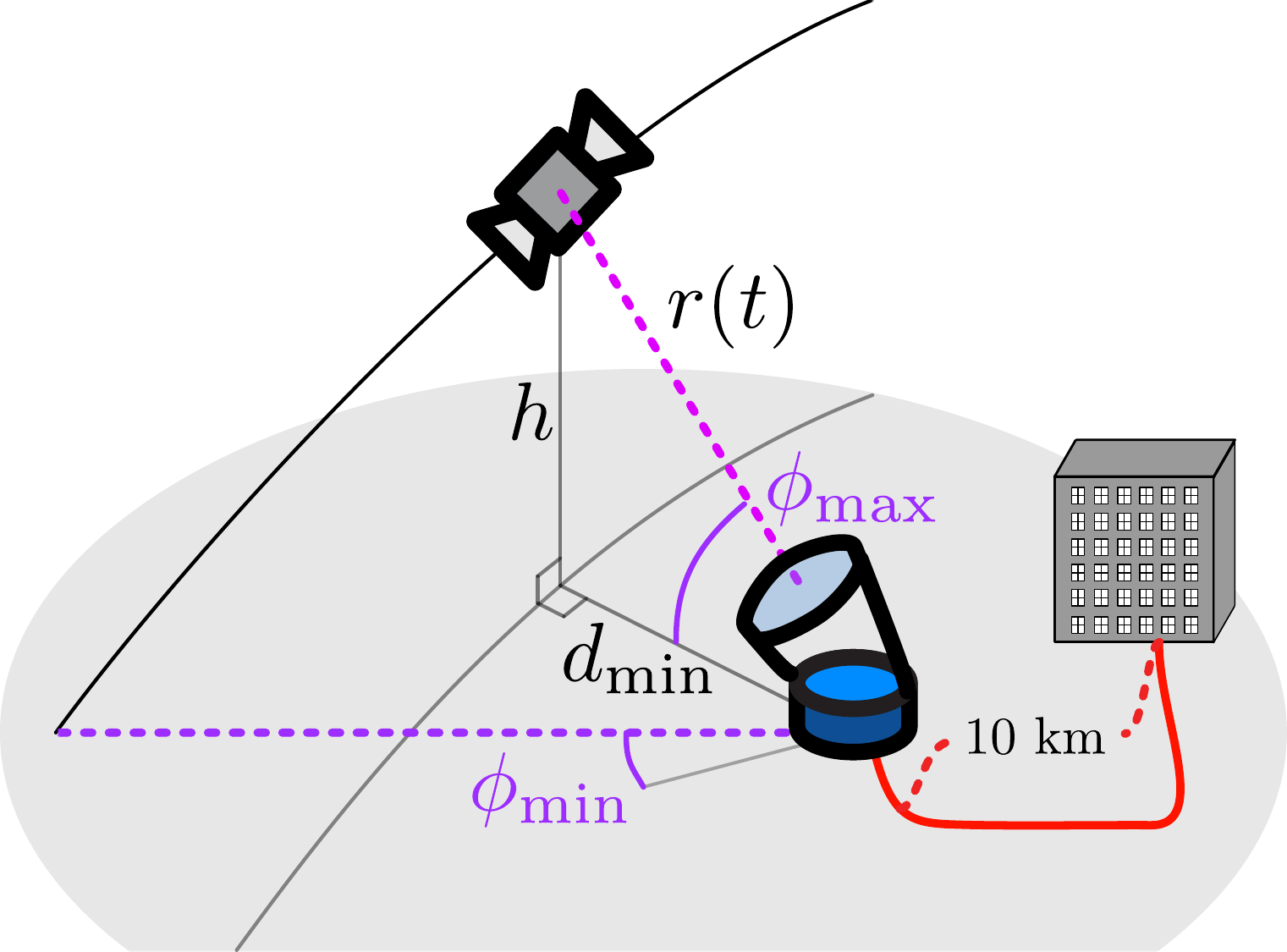}
    \caption{\textbf{Overpass geometry:} A schematic of a satellite overpass with altitude $h$, ground track offset $d$, and slant range $r(t)$. The maximum elevation angle $\phi_\text{max}$ is reached when the satellite is closest to the OGS, and the transmission is ended when the elevation angle dips under $\phi_\text{min}=15^\circ$. The EPS produces non-degenerate entangled photon pairs, and distributes them between the satellite and a metropolitan user, which is either 0 or 10~km away.}
    \label{fig:ogs_uplink}
\end{figure}%

\noindent
The scenario we consider is presented in Figure~\ref{fig:ogs_uplink}. The \emph{entangled photon source} (EPS), presented in Ref.~\cite{jaekenEmulationSatelliteUplink2025} is located at the OGS in Heriot-Watt and generates photon pairs encoded in a Bell state in polarisation. Each pair consists of a 1572~nm photon that is either detected immediately or routed through 10~km of single mode fibre SMF-28, and a 785~nm photon to be transmitted from a telescope to the satellite, while it is at an elevation angle $\phi$ with the horizon of at least $15^\circ$, dictated by the telescope design and the initial satellite acquisition procedure. Only a small fraction of the latter photons will arrive at the satellite receiver. This is quantified by the channel loss and can be attributed to different sources which we discuss individually here.

The total loss, $\smash{\eta_\text{total} = \eta_\text{atm}\eta_\text{bw}}$, is separated into contributions from atmospheric transmittance loss, $\eta_\text{atm}$ and beam widening $\eta_\text{bw}$ due to diffraction, turbulence and tracking errors, with
\begin{align}
    \eta_\text{bw} = \int_{A_\text{rec}} \frac{1}{2 \pi \sigma^2} \exp\left(-\frac{r^2}{2 \sigma^2}\right) dA,
\end{align}
where $A_\text{rec}$ is the receiving aperture area, $r$ is the radial distance from the beam centre, and $\sigma^2 = \sigma_{d}^2 + \sigma_{t}^2 + \sigma_{p}^2$ is the time-averaged beam size at the receiver, made up of independent contributions of diffraction ($\sigma_d$), turbulence ($\sigma_t$) and pointing and tracking error ($\sigma_{p}^2$).
Atmospheric transmittance loss at the transmission wavelength of 785~nm is modelled using libRadtran~\cite{libratran}, reproducing the result presented in Ref.~\cite{bourgoinComprehensiveDesignPerformance2013b}.

\newcommand*{\tabindent}{ \hspace{-1mm}}
\newlength{\thickarrayrulewidth}
\setlength{\thickarrayrulewidth}{2.1\arrayrulewidth}
\renewcommand{\arraystretch}{1.25}
\setlength{\tabcolsep}{8pt}
\begin{table}
    \centering
    \begin{tabular}{c|c}
         \thickhline
         \textbf{Parameter description} & \textbf{Value}\\
         \hline 
         Beam profile & Gaussian \\
         Diffraction ($\alpha_d$) & 25.5~$\mu$rad\\
         Pointing and tracking error ($\alpha_p$) & 3~$\mu$rad\\
         %\hline
         OGS aperture & 70~cm (HOGS)\\
         %\hline
         High-altitude wind speed ($v$) & 21~m~s$^{-1}$\\
         %\hline
         NG turbulence strength ($A$) & $1.7\times 10^{-14}\ $m$^{-\frac{2}{3}}$\\ \hline
         Receiver Aperture ($R_\mathsf{X}$) & 0.25~m \\
         OGS wavelength ($\lambda_\text{OGS}$) & 1572~nm \\
         Satellite wavelength ($\lambda_\text{sat}$) & 785~nm \\
         Beam spread at zenith ($\sigma_\text{zen}$) & 11.90~m\\
         Zenith transmittance ($\eta_\text{atm,zen}$) & 45.4\% \\ \hline

         Pump wavelength ($\lambda_p$) & 523.5~nm \\
         Maximum brightness ($B$) & 223~kHz~m/W  \\
         Heralding efficiency ($\eta_h$) & 19.3\% \\
         Entangled-state purity ($\mathcal{P}$) & 95.6±0.4\% \\
         Intrinsic errors (QBER, Qx) & 1.7\%, 3.9\% \\
         \thickhline
    \end{tabular}
    \caption{\textbf{Baseline parameters:} values for the simulation of the atmospheric and turbulence loss in a satellite up-link channel.
    The first set of parameters concern the ground station, specifically the \emph{Hub OGS} (HOGS) located in Edinburgh (UK).
    Turbulence conditions are also included, with the values for $A$ and $v$ being typical for sea level during nighttime~\cite{tofstedAtmosphericTurbulenceProfile2006}.
    The second set of parameters concern the satellite, specifically QEYSSat, and the beam properties incident on the satellite at an altitude of 550~km.
    These two sets of parameters are in line with the observed values in Ref.~\cite{renGroundtosatelliteQuantumTeleportation2017}.
    The final set of parameters concern the entangled-photon source itself.
    NG: Near-ground}
    \label{table}
\end{table}%

\begin{figure}
    \centering
    \includegraphics[width=\linewidth]{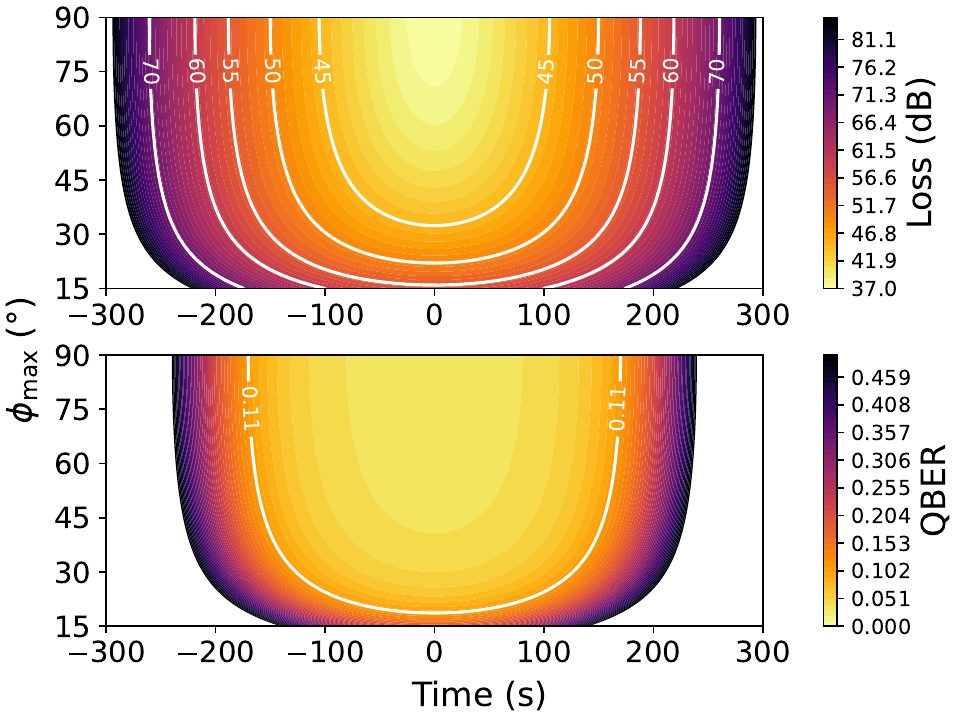}
    \caption{\textbf{Dynamic loss modelling:} (top) Up-link channel loss as a function of time during an overpass of a certain maximum elevation angle with respect to the OGS. (bottom) The corresponding instantaneous QBER 
    calculated based on experimental parameters obtained in Ref.~\cite{jaekenEmulationSatelliteUplink2025}.}
    \label{fig:loss_profiles}
\end{figure}

The contribution to the beam size due to diffraction is $\sigma_{d}(t)=\alpha_d/2 r(t)$, where $r(t)$ is the instantaneous channel length and $\alpha$ is the diffraction angle.
Similarly, the pointing and tracking error widens the beam as $\sigma_{p}(t)=\alpha_{p}/2 r(t)$. 
Atmospheric turbulence produces two distinct effects: beam expansion and beam wander. Beam expansion is caused predominantly by turbulent structures smaller than the beam diameter, which introduce spatially varying wavefront distortions and broaden the beam about its instantaneous centroid.
Conversely, turbulent structures larger than the beam diameter act approximately as refractive prisms, producing an overall wavefront tilt and displacement of the beam centroid, known as beam wander~\cite{mukherjee_analysis_2018,dios_scintillation_2004}.
In the present model, $\sigma_{t}$ describes turbulence-induced beam expansion through the Hufnagel-Valley refractive-index structure profile.
Beam wander is not included in the present model.
Beam expansion is modelled as follows,
\begin{align}
    \begin{split}
    \sigma_{t} = & \frac{2\sqrt{2}R(t)\lambda}{\pi} 
    \Big[1.46 \sec\left(\frac{\pi}{2}-\phi(t)\right)\left(\frac{2\pi}{\lambda}\right)^2 \\ & \int_0^h dz \; C^2_n(z)\left(1-\frac{z}{h}\right)^{\frac{5}{3}}\Big]^{\frac{3}{5}},
    \end{split}
\end{align}
where the satellite altitude, $h$, is assumed fixed to a circular orbit and $C_n^2(z)$ is the refractive-index structure constant:
\begin{align}
\begin{split}
    C_n^2(z) =~& 5.94\times10^{-53}\left(\frac{v}{27}\right)^2z^{10}e^{-\frac{z}{1000}}\\ &+ 2.7 \times 10^{-16} e^{-\frac{z}{1500}}+Ae^{-\frac{z}{100}}.
    \end{split}
\end{align}
The numerical values for all parameters can be found in Table~\ref{table}. The overpass is defined by $R(t)$ and $\phi(t)$ and can be parameterised by the maximum elevation resulting in loss profiles as shown in Figure~\ref{fig:loss_profiles}. The corresponding \emph{quantum bit error rate} (QBER) is based on the results of an experimental emulation~\cite{jaekenEmulationSatelliteUplink2025}. There, two experimental settings were considered where the 1572~nm photon was either detected immediately, or transmitted through 10~km of SMF-28 inducing dispersion of 18~ps/km/nm and a loss of 0.2~dB/km.

%

%--

\subsection{Secure key analysis of a satellite overpass}
\label{sec:secure_key_analysis}
 
\noindent
In quantifying the \emph{asymptotic key length} (AKL) of a QKD protocol, perfect knowledge of the error rate is assumed. For high-loss applications, such as satellite-based QKD, the error rate is calculated from the weighted mean of the instantaneous error rate
\begin{align}
    \overline{\text{QBER}}_{\text{key}} =\frac{\int_\mathcal{R} \; dt \; (\text{QBER}(t) \times \text{CC}(t))}{\int_\mathcal{R} \; dt \; \text{CC}(t)},
\label{eq:single_block}
\end{align}
where the region $\mathcal{R}$ is such that $\phi(t)\in[\phi_\text{min}, 180^{\circ}-\phi_\text{min}]$, $\text{QBER}(\phi)$ and $\text{CC}(\phi)$ are the instantaneous quantum bit error rate and coincidence click rate, respectively at time $t$, and $\smash{\phi_\text{min}}$ defines the minimum elevation angle over which data blocks are generated. The error rate in the complementary basis, denoted $\overline{\text{Qx}}$, is defined similarly. In the asymptotic limit of infinitely many detections, $\overline{\text{QBER}}_{\text{key}}$ represents the true average bit error probability. Note that Eq.~\eqref{eq:single_block} processes overpass data as a single block to which we apply finite statistics and uncertainties
to maintain high levels of composable security. This method scales more efficiently for large satellite networks since it removes the need to store and manage a multiple link segments until enough data is collected for key extraction. Section~\ref{subsec:data_postprocessing} explores alternative key postprocessing introducing multi-block strategies that combine overpass data with identical statistics to optimise key volume.

In practice, however, only a finite number of photon pairs are detected during a single overpass. The total sifted block size is $m=\frac{1}{2}\int_\mathcal{R} \; dt \; \text{CC}(t)$, where the factor $1/2$ accounts for basis sifting. From this block, a subset of size $k$ is randomly selected for parameter estimation, leaving $n=m-k$ signals for raw key generation. We therefore only have access to a small data sample to estimate the quantities $\text{QBER}(t)$ and $\text{Qx}(t)$ from the test sample mean of size $k$ bits. The sample mean $\overline{\text{QBER}}_{\text{sample}}$ is therefore a statistical estimate of the true quantity $\overline{\text{QBER}}_{\text{key}}$.

Since parameter estimation is performed on only a subset of the sifted data, these statistical fluctuations imply a non-zero probability that the observed sample error rate underestimates the true error rate affecting the remaining key bits. To formalise this we apply the sampling bound given by Lemma 1 of~\cite{Lim}. Namely, denoting the tolerated error rate threshold as $\delta$ and the finite-size buffer as $\nu$, the quantity
\begin{align}
    \epsilon_\text{pe}'(\nu,\xi)^2 = e^{-\frac{2mk\xi^2}{n+1}}+e^{-2((n\nu')^2-1)[\frac{1}{m(\delta+\xi)+1}+\frac{1}{m-m(\delta+\xi)+1}]}
\end{align}
bounds the failure probability that the protocol accepts (i.e. $\overline{\text{QBER}}_{\text{sample}} \leq \delta$) despite the actual key error rate exceeding the tolerated threshold by at least $\nu$ ($\overline{\text{QBER}}_{\text{key}} \geq \delta + \nu$), where $\nu'=\nu-\xi$, and $\xi \in (0, \nu)$ is an auxiliary deviation parameter chosen such that $n^2(\nu-\xi)^2>1$ and $m(\delta+\xi)\in \mathbb{Z}^+$. 
The first exponential term captures statistical fluctuations between the parameter-estimation sample and the remaining block, while the second term bounds the probability that the deviation exceeds the tolerated margin. 
The structure reflects concentration inequalities appropriate for random sampling without replacement, since the test bits are drawn from the same finite sifted block that later forms the raw key.

Analogously, the failure probability associated with privacy amplification quantifies the event that the final key is not sufficiently close to an ideal uniform key independent of an adversary. Privacy amplification applies a randomly chosen two-universal hash function to the raw key of length $n$, producing a final key of length $\text{SKL}$. The probability that this extracted key exceeds the available conditional smooth min-entropy of the raw key, in which case residual correlations with the adversary could remain, is then bounded by the quantity
\begin{align}
    \epsilon_\text{pa}(\nu) = \frac{1}{2}\sqrt{2^{-n(1-h_2(\delta+\nu)-f h_2(\delta))+t+\text{SKL}}},
\label{eqn:pa}
\end{align}
where $h_2$ is the binary entropy, $f$ the error correction efficiency, for which we use 1.19, as in Ref.~\cite{Lim}, and $t$ the length of the hash used for error correction. We now have all the ingredients to apply the composable security framework by Renner (e.g. see ~\cite{portmannCryptographicSecurityQuantum2014}). Namely, we can bound the trace distance between the joint final state of Alice, Bob and Eve and the corresponding ideal state, such that the final key is uniformly random, completely correlated between Alice and Bob and completely independent of Eve, by $10^{-s}$. We then say the protocol is secure to a parameter $s$, if there exists $0 < \xi < \nu < 1/2-\delta$ satisfying~\cite{muller-quadeComposabilityQuantumCryptography2009,portmannCryptographicSecurityQuantum2014}:
\begin{equation}
    10^{-(s+2)} +2\epsilon'_\text{pe}(\nu,\xi)+\epsilon_\text{pa}(\nu)\leq 10^{-s}.
\end{equation}
In accordance with Ref.~\cite{Lim}, we maintain a security level of $s=6$ as a realistic goalpost. 
Respecting these bounds, the SKL can be optimised over the internal parameters:
\begin{equation}
    \max_{\nu,\xi,k} \; \; \text{SKL},~~~~~~~~ k<m/2.
\end{equation}
In this work, we consider an experimental scenario based on experimental data, and use the estimate of $(\text{QBER}+\text{Qx})/2$ obtained from this as
$\delta$. 

%--

\section{Applications}
\label{sec:applications}

\noindent
Having established a physical model for the elevation-dependent up-link loss (section~\ref{sec:up-link}) and a finite-key optimisation framework for determining the achievable secure key length under statistical constraints (section~\ref{sec:secure_key_analysis}), we now apply these tools to assess the operational performance of up-link SatQKD.
In particular, our framework enables an evaluation of a number of mission-level design questions.
First, in section~\ref{subsec:optimal_power}, we extend the key rate optimisation to include the entangled photon source pump power, selecting an optimal value for each overpass, which remains constant during that pass
This overpass-specific optimisation is implemented for the remainder of the results.
Second, in section~\ref{subsec:dc_SKL} we determine how sensitive the key generation performance is to extraneous counts at the satellite receiver.
Finally, section~\ref{subsec:annual_SKL} explores how these effects combine over the distribution of overpasses throughout the year to determine the annual secure key yield.
Together, these studies translate the underlying physical and statistical modelling into quantitative guidance for mission design, including source operating parameters, detector requirements, and expected long-term key generation capability.
The computational codebase, optimisation routines, and dataset files generated for all applications considered are openly available on Zenodo~\cite{Sidhu_2026_software}.

%--

\subsection{Optimal pump power}
\label{subsec:optimal_power}

\noindent
For a given OGS, successive satellite overpasses occur with distinct maximum elevations and corresponding loss profiles as illustrated in Figure~\ref{fig:loss_profiles}. The pump power, $P$, that optimises the SKL will therefore differ for each overpass. Note that increasing the pump power directly increases the expected data blocks, but also increases the rate of accidental clicks in the same coincidence detection window and therefore the QBER. We consider a coincidence window of 400~ps for the case of no terrestrial fibre, and 1~ns in the case of 10~km of terrestrial fibre, as in Ref.~\cite{jaekenEmulationSatelliteUplink2025}. The pump power governs a non-trivial trade-off between signal rate and error rate, which heavily depends on the loss profile of each overpass.

An advantage of our up-link entanglement-based scheme is that the EPS is on the ground. This accessibility means the power can be %dynamically
tuned as required between overpasses. To inform the optimal pump power, we first include the pump power as a co-optimised parameter $\text{SKL}(P)$ to the finite key in section~\ref{sec:secure_key_analysis}, where the dependence on $P$ arises through the elevation-dependent coincidence rate and error statistics. The pass-dependent optimal pump power is given by
\begin{align}
P_\text{opt} = \argmax_{P\in\mathcal{P}} \text{SKL}(P)\, ,
\label{eqn:pump_power}
\end{align}
where $\mathcal{P}$ denotes the range of pump powers permitted by the source.

\begin{figure}[t!]
    \centering
    \includegraphics[width=\linewidth]{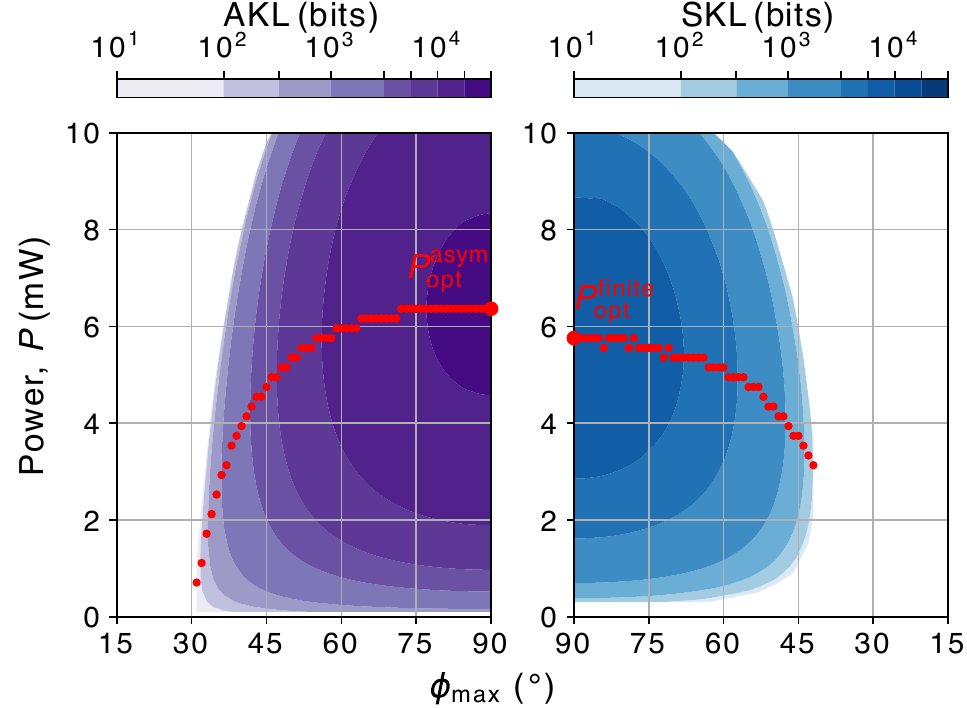}
        \caption{\textbf{Key lengths with pump powers and overpass geometries:} The asymptotic (left) and finite (right) key lengths with 10~km of terrestrial fibre are illustrated for a single overpass with a certain maximum elevation angle for different pump powers. These estimates are based on a satellite at an altitude of 550~km with an aperture of 25~cm, representative of the QEYSSat mission. The feasible region is smaller in the finite-statistics regime, and the optimal pump power for a zenith overpass is lowered to $P_\text{opt}^\text{finite}=5.76$~mW from $P_\text{opt}^\text{asym}=6.36$~mW. A corresponding figure with 0~km of fibre is shown in Supplementary Figure~\ref{fig:optim_0km}.
        }
    \label{fig:optim}
\end{figure}%

Fig.~\ref{fig:optim}(left) illustrates how the asymptotic key length (AKL) varies with different overpasses (characterised by the maximum elevation) and pump powers, with the corresponding finite SKL illustrated in Fig.~\ref{fig:optim}(right). In the asymptotic regime, the key rate $\smash{\text{AKL} = n(1-f h_2(\text{QBER})-h_2(\text{Qx}))}$ depends only on the true coincidence rate and the probability of error, without additional penalties from finite statistics. The optimal pump power, $\smash{P_\text{opt}^\text{asym}}$, is therefore set solely by the balance between increasing detected pair rate and the increase in QBER due to multi-pair induced accidental clicks. 

In the finite-key regime, the admissible pump-power range generating non-zero key is reduced relative to the asymptotic case. From Eq.~\ref{eqn:pa}, we see that $\text{SKL} \sim n (1 - h_2(\delta + \nu)-r) - \text{`\emph{finite correction terms}'}$. Heuristically, the reduced range is therefore due to the additional safety margin $\nu$ and the exponentially growing penalties in parameter estimation and privacy amplification. For the same reason, the optimal pump power is lowered for a fixed $\phi_{\text{max}}$, including for a zenith overpass: $\smash{P_\text{opt}^\text{finite} < P_\text{opt}^\text{asym}}$.

%--

\subsection{Extraneous count sensitivity}
\label{subsec:dc_SKL}

\noindent
The satellite receiver inevitably registers detection events that do not originate from the EPS.
These arise from background photons within the receiver field of view and from intrinsic detector dark counts triggered by thermal excitations, both leading to extraneous counts.
When such spurious detections fall within the coincidence window of a ground-based detection, the observed QBER increases since these accidental coincidences are not correlated and contribute random outcomes. In this section, we quantify the sensitivity of the SKL to different \emph{extraneous count} (EC) rates to inform design choices of the detector technology, spatial and spectral filtering requirements, and operational constraints such as allowable background conditions.

The EC level may differ for each satellite overpass, depending on: \textbf{(i)} ambient light environment at the OGS site location, \textbf{(ii)} the timing of the overpass, and \textbf{(iii)} the detector operating temperature, which can vary with time. We therefore explore a range of EC rates to accurately model the up-link key generation performance. Our optimisation over the pump power in Eq.~\eqref{eqn:pump_power} enables operational flexibility in tuning the pump power to best suit the background conditions. Using experimentally characterised EC-dependent QBER data from Ref.~\cite{jaekenEmulationSatelliteUplink2025}, we interpolate the relationship between varying EC levels and the corresponding error probability and propagate this to our finite-key optimisation framework.

\begin{figure}
    \centering
    \includegraphics[width=\linewidth]{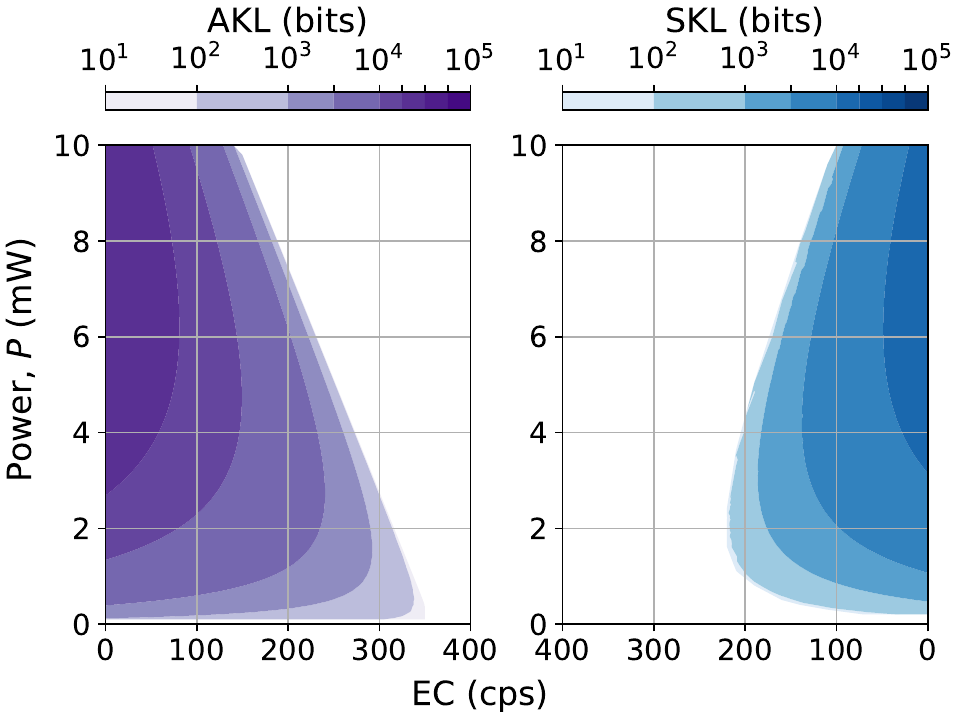}
    \caption{\textbf{Key lengths with pump powers and extraneous count rates:} the asymptotic (left) and finite (right) key lengths with 10~km of terrestrial fibre are illustrated for a single zenith overpass with a varying level of satellite detector EC and pump powers. The ground detectors have a fixed EC of 200~cps. These estimates are based on a satellite at an altitude of 550~km with an aperture of 25~cm, representative of the QEYSSat mission. The feasible region is smaller in the finite-statistics regime as would be expected, and the optimal pump power is decreased for low EC, but increased for the high EC. A corresponding figure with 0~km of fibre is shown in Supplementary Figure~\ref{fig:DC_0km}.}
    \label{fig:DC}
\end{figure}

Figure~\ref{fig:DC} illustrates the resulting SKL for a zenith overpass as a function of pump power and satellite detector EC levels, where two clear trends emerge.
First, increasing the extraneous count rate reduces the achievable SKL across all pump powers and narrows the feasible operating region.
This occurs because extraneous counts contribute a baseline of uncorrelated detection events, directly increasing the QBER.
Second, as the EC rate increases, the optimal pump power decreases in both the asymptotic and finite-key regimes.
Specifically, an increasing baseline QBER reduces the amount of power-induced (multi-pair) errors that can be tolerated without exceeding the security threshold. 
In the scenario with 10~km of terrestrial fibre, the maximal tolerable EC in the satellite is 350~cps in the asymptotic regime, and 215~cps in the finite-key regime.
Without the terrestrial fibre these are 1250~cps and 960~cps respectively, shown in Supplementary Figure~\ref{fig:DC_0km}.
This is realistic for an OGS in a rural location~\cite{yastremskiEstimatingImpactLight2025}, especially given recent developments in detector technology~\cite{yangSpaceborneLownoiseSinglephoton2019} and extraneous count mitigation strategies~\cite{yangInorbitDarkCount2024}.
For both scenarios, the ground detector EC level was set at 200~cps.

These results highlight two implications for mission design. First, detector extraneous count rate is a critical system parameter, directly determining not only peak SKL but also the admissible pump-power range. Second, the per-overpass dynamic adjustment of source brightness offers a practical mitigation strategy under varying background conditions, although it cannot fully compensate for excessive EC levels. Detector technology, filtering performance, and brightness control must therefore be co-designed to maximise long-term secure key yield.

%

%--

\subsection{Annual key rate}
\label{subsec:annual_SKL}

\noindent
The annual SKL provides a system-level metric for quantifying the long-term performance of a SatQKD link between a defined OGS location by averaging over all accessible overpasses. While this figure of merit has been used to assess prepare-and-measure~\cite{sidhu2021key, Sidhu2022_npjQI, Sidhu2023finite} and entanglement-based~\cite{Sidhu2026operational} downlink architectures, its application to up-link geometries remains unexplored. In this section, we address this to evaluate the long-term performance under realistic operating conditions that reflect both the QEYSSat and Micius missions. The secure key obtainable in a given overpass is highly dependent on the satellite's maximum elevation angle, which changes dramatically each overpass. In contrast, the annual SKL captures the combined effects of orbital geometry and channel loss, providing a realistic benchmark for mission design and expected key generation capacity.

\begin{figure}
    \centering
    \includegraphics[width=\linewidth]{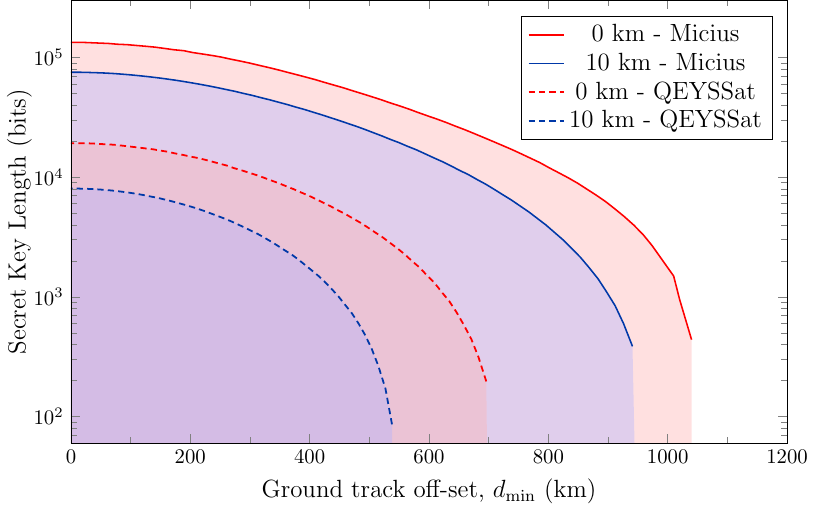}
    \caption{\textbf{Optimised SKL for different overpasses,} parameterised by different ground track offset values, $d_\text{min}$ (bottom axis), or equivalently the maximum elevation angle, $\phi_\text{max}$ (top axis). The SKL here is optimised over the pump power and all protocol parameters. Solid lines correspond to no optical fibre and dashed lines to a 10~km optical fibre length at the OGS. The representative scenarios model the QEYSSat mission with $h=550$~km and $R_\mathsf{X}=25$~cm and the Micius mission with $h=500$~km and $R_\mathsf{X}=30$~cm.}
    \label{fig:gto_integral}
\end{figure}

To evaluate the annual SKL, we consider all overpasses accessible to a given OGS, parameterised by their maximum elevation angle $\smash{\phi_\text{max} \in [\phi_\text{min}, 90^\circ]}$ as measured from the local horizon. For non-zenith passes, the orbital geometry can be equivalently described by the angle $\theta$, defined from the centre of the Earth and corresponding to the rotation required to map the overpass into an equivalent zenith trajectory. The relation between these quantities follows from the Earth-satellite geometry (see Supplementary Materials section~\ref{supp1} for details),
\begin{align}
\sin(\theta)\tan(\phi) + \frac{R}{R+h} = \cos(\theta),
\label{eq:earth}
\end{align}
yielding
\begin{align}
\theta(\phi) = \arccos\left[\frac{R}{R+h}\cos(\phi)\right] - \phi.
\label{eq:theta_eq}
\end{align}
Eq.~\eqref{eq:theta_eq} provides a natural parametrisation of overpasses in terms of ground-track offset, $d_\text{min}=R\theta$ (see Fig.~\ref{fig:ogs_uplink}), and is more convenient to determine the annual SKL. Figure~\ref{fig:gto_integral} illustrates the per-overpass optimised SKL as a function of ground-track offsets with and without terrestrial fibre at the OGS. As expected, the SKL is largest for zenith overpasses ($d_\text{min}=0$, $\phi_\text{max}=0^\circ$). As the ground-track offset increases, the satellite remains at lower elevations throughout the pass, resulting in an increased average link QBER and smaller block sizes. These effects compound to a regime where the interplay between channel loss and finite-key effects rapidly degrades key generation and the range of viable overpasses. Our simulations inform the maximum tolerable ground track offset with a 10~km fibre link and without fibre link as $\sim$538~km and $\sim$941~km respectively.

Assuming a circular polar orbit and neglecting Earth's rotation during a single overpass, successive orbital longitudes are treated as uniformly distributed. This allows the annual SKL to be expressed as
\begin{align}\label{eq:skl}
    \text{SKL}_\text{annual} = \frac{N_\text{orbits}^\text{year}}{L_\text{lat}} \int_{-r\theta(\phi_{\text{min}})}^{r\theta(\phi_{\text{min}})} \; dd_\text{min} \; \text{SKL}_{\text{opt}}(d_\text{min}),
\end{align}
where $d_\text{min}$ is the ground-track offset defining the minimum approach of the ground track to the OGS, $N_\text{orbits}^\text{year}$ is the number of orbits per year, $L_\text{lat}$ the longitudinal circumference along the line of latitude at the OGS site location, and $\smash{\text{SKL}_{\text{opt}}(d_\text{min})}$ the per-overpass optimised SKL. 
We assume that the transmission window is symmetric about the point of closest approach (local zenith), labelled $t=0$ ($\phi=90^\circ$), such that the total key generated per overpass is obtained by integrating over this symmetric interval. Due to this symmetry, the integrand in Eq.~\eqref{eq:skl} is twice the integrated (shaded) area in Fig.~\ref{fig:gto_integral}. In practice, the limits in Eq.~\eqref{eq:skl} can be extended to cover the whole latitudinal circumference, as the SKL is zero over most of the domain.

\begin{table*}[t!]
    \centering
\begin{tabularx}{\linewidth}{Y|Y|Y|Y|Y|Y|Y}
    \thickhline
    \multirow[c]{2}{*}{\centering \textbf{Mission}} 
    & \multirow[c]{2}{*}{\centering\arraybackslash\parbox{1.9cm}{\centering \textbf{Satellite altitude}}} 
    & \multirow[c]{2}{*}{\centering\arraybackslash\parbox{1.6cm}{\centering \textbf{Aperture}}} 
    & \multicolumn{2}{c|}{\textbf{Shot-by-shot Annual Key}} 
    & \multicolumn{2}{c}{\textbf{Integrated Annual Key}} \\
    \cline{4-7}
    &
    &  
    & \textbf{10~km}
    & \textbf{0~km}
    & \textbf{10~km}
    & \textbf{0~km} \\
    \hline
    $^\dagger$QEYSSat & 550~km $^{\dagger}$& 0.25~m $^{\dagger}$ & 0.91~Mb & 8.7~Mb & 2.3~Mb & 13.2~Mb\\
    
    $^\star$Micius & 500~km$^{\star}$& 0.3~m$^{\star}$ & 1.93~Mb & 14.5~Mb & 3.98~Mb & 20.4~Mb \\
    
    --- & 550~km$^{\dagger}$& 0.3~m$^{\star}$ & 1.76~Mb & 14.0~Mb & 3.78~Mb &  20.0~Mb \\
    
    --- & 500~km$^{\star}$& 0.25~m$^{\dagger}$ & 1.04~Mb & 9.12~Mb & 2.45~Mb &  13.6~Mb \\
    \thickhline
\end{tabularx}
\caption{\textbf{Annual SKL for representative satellite configurations:} The annual secure key length is shown for satellite altitudes and receiver apertures.
The first two rows are the parameters corresponding to the $^\dagger$QEYSSat and $^\star$Micius missions, while the following two rows are cross-combinations of these parameters.
Results are presented for both shot-by-shot and integrated annual keys, with different terrestrial fibre lengths at the OGS.
Aggregating detection events across all overpasses prior to key distillation yields significant improvements. The pump power is remains constant during each overpass, but differs on subsequent passes due to our dynamic optimisation.}
\label{tab:ann_key}
\end{table*}%

For our representative up-link configuration, we consider the \emph{Hub Optical Ground Station} (HOGS) in Edinburgh, UK (latitude $55.91^\circ$) with system parameters consistent with the upcoming QEYSSat mission implementing the BBM92 up-link configuration with $h=550$~km and $R_\mathsf{X}=25$~cm.
Since the QEYSSat satellite will be in a Sun Synchronous Orbit that is similar to a Polar Orbit, we can use Eq.~\eqref{eq:skl} as an approximation to the annual SKL.
At the HOGS latitude and satellite altitude, we find $\smash{N_\text{orbits}^\text{year}=5504}$ and the latitudinal circumference for HOGS is $L_\text{lat}=11,231$~km, which together set the normalisation in Eq.~\eqref{eq:skl}.
Table~\ref{tab:ann_key} summarises the resulting annual SKL for our representative link with HOGS, alongside the Micius mission ($h=500$~km, $R_\mathsf{X}=30$~cm, $\smash{N_\text{orbits}^\text{year}=5564}$) for comparison.
With QEYSSat's altitude currently intended within 500-600~km~\cite{hrynykCanadianSpaceAgency,jenneweinQEYSSATMissionProposal2014b}, additional rows correspond to cross-combinations of altitude and apertures to isolate parameter dependencies relevant to system design. 

Table~\ref{tab:ann_key} also distinguishes between two operational regimes: \textbf{(i)} in the shot-by-shot approach, a secure key is extracted independently from each overpass and subsequently summed, whereas \textbf{(ii)} in the integrated approach, detection events from all overpasses are aggregated prior to key distillation, suppressing finite key penalties.
In both cases, we consider configurations with and without 10~km of terrestrial fibre at the HOGS facility.
The quoted annual key sizes adopt an optimised cut-off minimal elevation angle, which is explained in more depth in Section~\ref{supp3} and Figure~\ref{fig:cutoff}.
As expected, increasing the aperture or reducing the orbital altitude improves the key yield by enhancing link efficiency across all configurations.
Further, the presence of fibre reduces the achievable key by an order of magnitude in all cases, underscoring the strong sensitivity to additional receiver-side loss.
Comparing the two operational regimes, we see that the integrated strategy significantly enhances the annual SKL, leading to a factor improvement of approximately 1.5 without fibre, and approximately 2.3 with 10~km of fibre.
As a general rule, the lower the annual key rate is to start with, the more of a difference this strategy makes.

We restrict the analysis to a single OGS location, as site placement in a global network is primarily driven by coverage constraints rather than local optimisation. These results demonstrate that both hardware choices and operational strategy critically impact long-term key generation. In particular, minimising loss and exploiting data aggregation across passes can yield an order-of-magnitude gain, providing a clear pathway for optimising future up-link SatQKD links for quantum networking applications.

%--

\section{Guide for implementation}
\label{sec:implementation}

\noindent
Our analysis establishes the optimised performance for entanglement-based satellite up-link quantum communication under realistic atmospheric, orbital, and detector conditions. The resulting secure key lengths and long-term network utility provide operational benchmarks for the underlying quantum network and its capability to reliably distribute entanglement across global distances. A natural question then follows: \emph{how can these performance regimes be realised in practice?}

There are a number of targeted answers to this questions. On hardware, the simulations place concrete requirements on the entangled-photon source and detection system, many of which are already accessible with current photonic and detector technologies, particularly in architectures where the entangled source remains ground based. Further, improvements are possible through protocol-level refinements that are tailored to finite block lengths, and adaptive post-processing strategies that exploit the strongly time-varying quality of satellite overpasses. In the following sections, we examine the hardware requirements and technological readiness (Section~\ref{subsec:hardware}), and post-processing strategies (Section~\ref{subsec:data_postprocessing}) required to realise and enhance the performance of practical up-link quantum networking.

%--

\subsection{Hardware}
\label{subsec:hardware}

\noindent
On hardware, achieving the simulated performance regimes requires high-brightness entangled-photon generation, efficient photon collection, low-noise detection, and precise timing synchronisation. The entangled-photon source demonstrated in our previous experimental study already operates close to this required regime, achieving a measured brightness of 223,000 pairs/s/mW, a symmetric heralding efficiency of 19.3\%, a state purity of 95.6\%, and QBER = 1.7\% and Qx = 3.9\% giving an entanglement fidelity of 94.4\%.

As illustrated in Fig.~\ref{fig:optim}, the optimal pump power lies well within the experimentally characterised range of the source.
Increasing the pump power beyond this optimum produces more multi-pair and accidental coincidences, thereby increasing the error rate.
Consequently, it is more important to have a high degree of dynamic control of the pump power than maximising brightness alone, as it allows the source to be optimised on a per-overpass basis.
This is particularly advantageous in the up-link architecture considered here, where the source remains ground based and can therefore be adaptively reconfigured between overpasses.
Our simulations (Fig.~\ref{fig:optim}) specifically indicates an increase of several mWs of pump power and yields an order of magnitude improvement in key servicing.
Improving heralding efficiency is often a more effective route to extending the operating range of a photon-pair link than simply increasing source brightness.
Higher heralding efficiency increases the probability that a generated photon pair produces a detected coincidence, whereas increasing brightness alone eventually increases multi-pair emissions and accidental coincidences, degrading QKD performance.
For representative operating points, doubling the heralding efficiency increases the tolerable channel loss by $\sim$4~dB, thereby extending the achievable secure operating regime and increasing the attainable key length. Source development should therefore prioritise collection efficiency, spatial mode matching, and focusing conditions in order to maximise heralding efficiency and overall network performance.

Detector performance is similarly critical to attain the reported finite-key performances. Detectors aboard the satellite must maintain sufficiently low extraneous count rates, since uncorrelated detection events generate accidental coincidences that directly increase the QBER. As illustrated in Fig.~\ref{fig:DC}, finite-key operation with satellites is substantially more sensitive to detector noise than is revealed with asymptotic operation. In practical deployment, background photons arise from urban illumination, atmospheric scattering, and reflected moonlight~\cite{bourgoinComprehensiveDesignPerformance2013b,bonato_feasibility_2009}. Their effects become increasingly pronounced in low-elevation passes, where the link loss is highest and the finite-key operating regime is most constrained.

Active suppression of detector noise is easily achieved through spectral filtering and cooling.
Superconducting nanowire single-photon detectors (SNSPDs) offer state of the art performance, achieving high detection efficiencies, low extraneous counts, and low timing jitters.
Our previous work utilised SNSPDs to capture the 1572~nm photons, with each detector having a nominal detection efficiency of 80\%, dark count rate of up to 100~cps, jitter of 50~ps, and a dead time of 25~ns, with negligible background photons.
However, SNSPDs require cryogenic operation and typically rely on single mode coupling, presenting significant challenges for deployment aboard satellites.
Despite these challenges, there are ongoing efforts to deploy SNSPDs for space applications~\cite{you_superconducting_2018,liu_development_2021}, with a particular focus on the development of space-qualifiable cryogenic cooling systems~\cite{tan_1_2024}. 
By contrast, silicon single-photon avalanche photodiodes (Si-SPADs) present a more practical near-term solution for satellite deployment.
For example, an Excelitas SPCM-AQRH-15-FC SPAD typically exhibits a quantum efficiency of 60\%, timing jitter on the order of $\sim$0.5~ns, dark count rate $<$50~cps, and a maximum dead time of 30~ns.
From Fig.~\ref{fig:DC}(right), this operating regime remains compatible with key generation of $\sim$\num{9.88e3}~bits at a pump power of 6.16~mW.
However, emerging detector platforms with higher quantum efficiency and reduced dark count rates of $<$25~cps~\cite{an_silicon_2025} would offer near-term enhancement of secure key throughput, particularly in high-loss or fibre-integrated deployment scenarios.
For a detector with a dark count rate of 25~cps, key generation increases to \num{12.48e3}~bits at a pump power of 6.67~mW, showing a performance increase of 26.3\%.
Additional considerations arise from radiation damage, which has been shown in previous satellite missions to increase dark count rates over operational lifetimes~ \cite{yangInorbitDarkCount2024,yangSpaceborneLownoiseSinglephoton2019,lenart2025comparing}. Timing performance also directly constrains the minimum achievable coincidence window.
Narrow coincidence windows suppress accidental coincidences and background-induced QBER, but can also reject true coincidence events once the coincidence window approaches the combined detector and synchronisation jitter.
For SPADs such as the Excelitas, a coincidence window of $\sim$1~ns would be appropriate, as was the case in our previous work for a 10~km terrestrial fibre link.
Achieving the simulated operating regimes therefore requires low-jitter detectors, high-stability time-tagging electronics, and synchronisation sufficiently precise to support sub-nanosecond coincidence windows.
These requirements become particularly important in real-world deployments involving kilometre-scale terrestrial fibre links, where chromatic dispersion and synchronisation drift can significantly broaden the coincidence distribution.

%--

\subsection{Data post-processing}
\label{subsec:data_postprocessing}

\noindent
In this work, we optimise the finite SKL obtained from a single block received during a satellite overpass. The low-elevation regions at the beginning and end of the overpass experience higher loss and contribute fewer signals. Including this data therefore increases the average overpass QBER, reducing the optimised SKL. To mitigate this effect, we applied an optimised cut-off elevation angle for the results presented in Sec.~\ref{sec:applications}, effectively addressing the trade-off between the single-block size and the average overpass QBER. 

We consider a second strategy of segmenting the overpass data into multiple blocks to try and increase the SKL. Intuitively, this improvement would arise from data that would otherwise be discarded, but which may be independently processed to distil non-zero keys. In the asymptotic regime, dividing a block into smaller sub-blocks can be advantageous when the errors are not uniformly distributed. Specifically, owing to the concavity of the binary entropy function, processing data with different error rates separately can increase the total secret key extracted relative to processing the same data as a single block. Recent work has suggested that similar ideas may also be beneficial in the finite-key regime~\cite{baeBlockwisePostprocessingSatellitebased2025, Wang2026scalability}. To develop intuition and quantify the advantage of alternative data post-processing strategies for our up-link geometry, we consider the simplest non-trivial case and divide the data into two blocks. 

Consider segmenting a raw block of size $m$ and average error rate $\text{QBER}$ into two sub-blocks of sizes $m_1$ and $m_2 = m - m_1$, with corresponding error rates constrained by
\begin{align}
m_1 \text{QBER}_1 + m_2 \text{QBER}_2 = m \text{QBER} \, .
\label{eq:qber_restriction}
\end{align}
This framework naturally extends to a larger number of blocks and is agnostic to the physical origin of the data. Further, Eq.~\eqref{eq:qber_restriction} preserves both the total block size and average QBER while allowing the ratio $m_1/m_2$ and the distribution of errors between the two blocks to vary. Any change in SKL therefore arises solely from how the errors are distributed within the data.

\begin{figure}[t!]
    \centering
    \includegraphics[width=\linewidth]{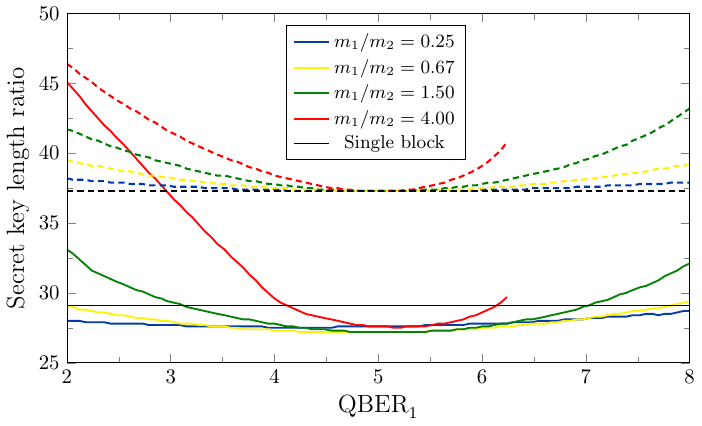}
    \caption{\textbf{Key length for two block post-processing:} SKL, normalised to the raw block size, for a block of $m=10^6$ and average $\text{QBER}=5\%$ as a function of sub-block sizes $m_1$ and $m_2 = m - m_1$. For each fixed ratio, $m_1/m_2$, we vary $\text{QBER}_1$ with $\text{QBER}_2$ determined from Eq.~\eqref{eq:qber_restriction} to ensure the total block size and average QBER remain unchanged. The asymptotic and finite-key regimes are represented by dashed and solid lines receptively. In both cases, the black line illustrates the SKL without block segmentation.}
    \label{fig:multiblock}
\end{figure}

Fig.~\ref{fig:multiblock} illustrates the raw block size normalised SKL ratio for different block ratios $m_1/m_2$ as a function of the $\text{QBER}_1$, with $m=10^6$ and $\text{QBER}=5\%$. The dashed curves show the asymptotic case. As expected, separating overpass data never performs worse than processing it as a single block, and can provide a modest improvement when one block has a lower error rate than the average. The finite-key regime (solid lines) exhibits markedly different behaviour. Dividing the data generally reduces the achievable SKL because each block incurs an independent finite-size penalty. Improvements are observed only within a restricted parameter space corresponding to $m_1 \gg m_2$ and $\text{QBER}_1 \ll \text{QBER}$, so that the majority of the data reside in a large, low-error block.

Operationally, any implementation whose raw key exhibits an error distribution within this favourable region of parameter space can benefit from two-block post-processing. For SatQKD, the highest-elevation portion of an overpass contributes most of the detected photons and therefore dominates the average QBER. Consequently, the practically accessible operating regime is typically characterised by $\text{QBER}_1 \approx \text{QBER}$, rather than the strongly asymmetric regime where a two-block advantage is observed. In this case, splitting the overpass data generally reduces the achievable finite-key rate. For the applications considered in this work, the most effective strategy is therefore to maximise the size of the high-quality block $m_1$, often leaving the remaining block $m_2$ too small to generate a positive key due to finite-size penalties. This strategy is equivalent to discarding the highest-error portion of the overpass, as implemented through an optimised cut-off elevation angle in Sec.~\ref{sec:applications}.

While Fig.~\ref{fig:multiblock} demonstrates that there may exist scenarios where retaining and processing a second block can yield additional secret key, our results suggest that for the up-link scenarios studied here, this advantage is not accessible.
Whether more sophisticated multi-block partitioning schemes can provide additional improvements remains an open question and is a promising direction for future work, which could be addressed by directly optimising the block boundaries using the time-dependent error and detection profiles of a satellite overpass to determine whether additional gains in secure key length are achievable. 

%--

\section{Discussion}
\label{sec:conclusions}

\noindent
The maturity and scalability of distributed quantum technologies rest on the distribution and manipulation of quantum resources over global scales. Within this vision, the satellite up-link configuration provides a fundamental primitive for scalable quantum networking. A satellite equipped to perform delayed Bell-state measurements can entangle photons arriving from two widely separated OGSs, thereby acting as a trusted, reconfigurable network node that interconnects otherwise isolated metropolitan quantum networks. From an engineering perspective, this up-link geometry also anchors complex quantum source and associated hardware on the ground, permitting modular upgrades that directly facilitate the rapid adoption of experimental advances in these technologies. Alternatively, an uplink scenario is also conceptually interesting because it allows for a measurement-device-independent setting \cite{lo2012measurement,braunstein2012side} where two ground stations with trusted sources establish a link with an untrusted satellite that performs a joint measurement of the two incoming signals. This might be advantageous for a variety of reasons where the satellite would be deemed untrusted. Returning to our original motivation of providing the first quantitative feasibility analysis of up-link satellite quantum communications under rigorous finite-key security analysis, this work addresses a critical gap that has previously left the operational viability of the up-link configuration an open question.

Our results demonstrate that entanglement-based up-link SatQKD remains robustly feasible even in the finite-key regime. By coupling a rigorous finite-key security analysis with realistic orbital dynamics representative of the imminent QEYSSat mission, we evaluate the optimised SKL performance using the measured hardware parameters of the UK Quantum Communications Hub OGS at Heriot-Watt University. Our analysis shows that considering finite statistics demands a shift in the optimal operating regime relative to asymptotic predictions, requiring system parameters to be re-optimised for practical operation. In particular, modest reductions in source brightness provide an accordant mitigation of finite-size penalties by suppressing the increase in quantum bit error rate associated with multi-pair emission. The optimal pump power is approximately 10\% lower for overpasses close to zenith and up to 30\% lower for overpasses reaching a maximum elevation of 45$^\circ$, leading to an increase in the extractable secure key length.
Another main finding of our analysis is that, in comparison with the asymptotic analysis, the tolerance of the protocol to background counts is reduced significantly, to about 61\% of its maximum value when the ground-based user is 10~km removed from the OGS and 77\% when this user is co-located with the OGS.

Beyond optimising individual overpasses, a relevant metric for a practical satellite mission is the cumulative secure key generated across the full distribution of accessible orbital passes. Across a complete annual cycle, we estimate secure key yields of approximately 13~Mb for a QEYSSat-like mission and 20~Mb for a Micius-like mission (integrated annial key with no optical fibre in Table~\ref{tab:ann_key}), with modest variations arising from differences in satellite altitudes and receiver aperture. For comparison, an equivalent down-link architecture is expected to achieve approximately an order of magnitude higher secret key yields because of its substantially lower channel losses~\cite{Sidhu2026operational}. To mitigate this finite-size bottleneck, we show that the accumulation of raw key material from multiple overpasses before key extraction substantially increases the annual secure key yield. However, this improvement comes at the expense of increased latency and storage requirements, which makes its practical adoption dependent on the requirements of the target application.

Our results also clarify the role of blockwise post-processing in finite-key SatQKD. While asymptotic analyses predict that partitioning overpass data with different error characteristics can increase the extractable SKL, our finite-key analysis reveals that the statistical penalties associated with independently estimating the security parameters for each block generally outweigh this advantage. Although a narrow region of parameter space exists in which two-block processing improves the secure key length, the characteristic error profile of a satellite overpass places realistic SatQKD operation outside this favourable regime. Consequently, the preferred strategy is to retain the largest high-quality data block while discarding the highest-error portion of the overpass through an optimised elevation cut-off. These findings suggest that simple blockwise post-processing is unlikely to provide further gains for near-term SatQKD missions, but do not preclude improvements from more sophisticated multi-block strategies, whose optimal boundary configurations remain an important direction for future investigation~\cite{wangScalabilityBottlenecksEntanglementbased2026,baeBlockwisePostprocessingSatellitebased2025}.
Further refinements to the predictive framework may also be achieved by incorporating adaptive error-correction efficiencies that depend explicitly on the observed QBER, block size and security parameters~\cite{tomamichelFundamentalFiniteKey2017a}.

%--

\section*{Acknowledgements}

\noindent
This work was supported by the UK Engineering and Physical Sciences Research Council (Grant Nos. EP/Z533208/1, EP/W027011/1, EP/Z533956/1), the Horizon Marie Sk{\l}odowska-Curie Fellowship (Project number: 101211639), the Innovate UK project ReFQ (Project number: 78161), and the BMFTR project QuantenRepeater.Net (QR.N). We thank Jens Eisert and Nathan Walk for fruitful discussions. 

%--

%\bibliography{bib.bib}
\bibliographystyle{naturemag}

%--

\clearpage
\setcounter{figure}{0} 
\onecolumngrid

\section*{Supplementary Material}
\label{sec:supp}

\noindent
Here we provide supplementary material about the results in our paper. Section~\ref{supp1} provides additional results equivalent to those in Figures~\ref{fig:DC} and~\ref{fig:optim}, with the difference that the 10~km of terrestrial fibre is omitted. Section~\ref{supp1} provides a graphic explanation of how the relation between the ground-track offset and the elevation angle may be obtained.
\renewcommand{\thefigure}{S\arabic{figure}}
\setcounter{figure}{0}

%--

\subsection{Contour plots with 0~km of fibre}
\label{supp12}

\begin{figure}[h!]
    \centering
    \includegraphics[width=0.6\linewidth]{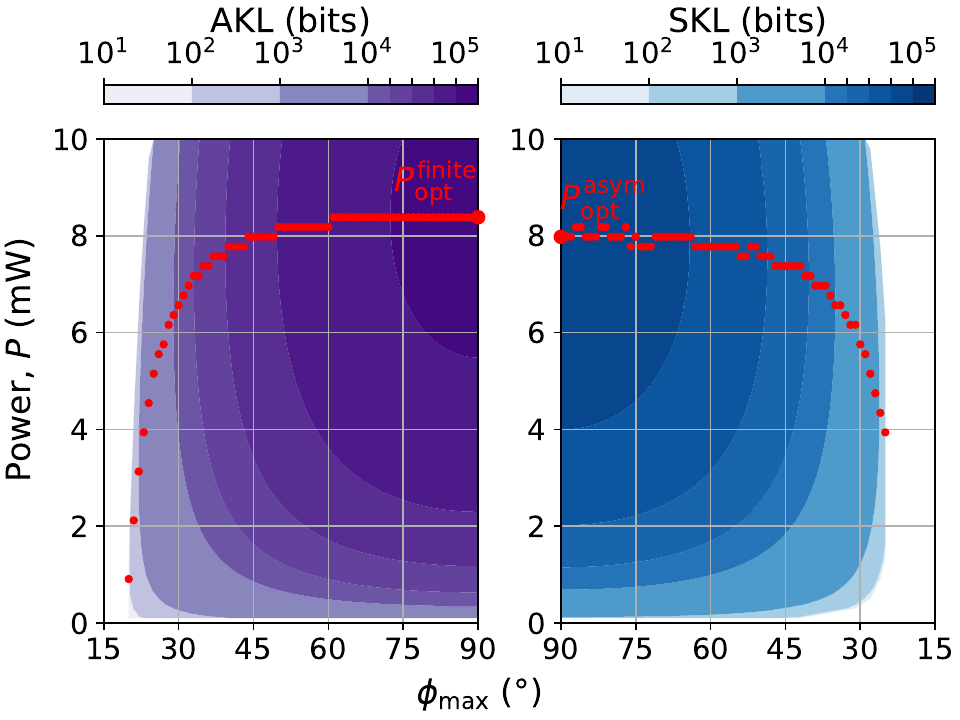}
        \caption{\textbf{Key lengths with pump powers and overpass geometries:} The asymptotic (left) and finite (right) key lengths with 0~km of terrestrial fibre are illustrated for a single overpass with a certain maximum elevation angle for different pump powers. These estimates are based on a satellite at an altitude of 550~km with an aperture of 25~cm, representative of the QEYSSat mission.}
    \label{fig:optim_0km}
\end{figure}

\begin{figure}[H]
    \centering
    \includegraphics[width=0.6\linewidth]{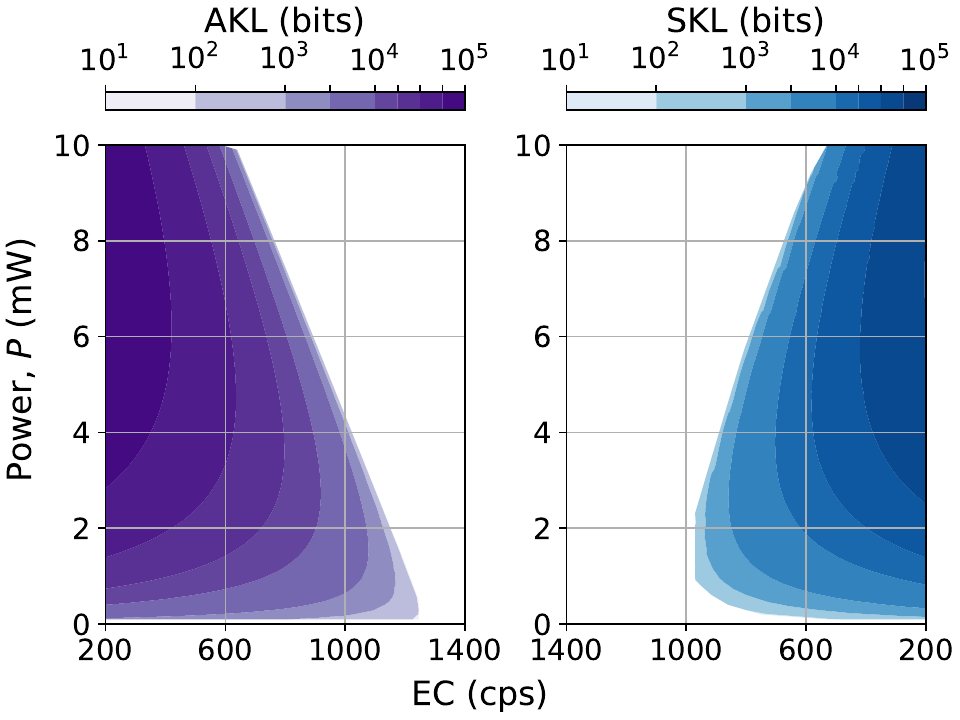}
    \caption{\textbf{Key lengths with pump powers and extraneous count rates:} the asymptotic (left) and finite (right) key lengths with 0~km of terrestrial fibre are illustrated for a single zenith overpass with a varying level of satellite detector EC and pump powers. The ground detectors have a fixed EC of 200~cps. These estimates are based on a satellite at an altitude of 550~km with an aperture of 25~cm, representative of the QEYSSat mission.}
    \label{fig:DC_0km}
\end{figure}

\noindent
Figures~\ref{fig:optim_0km} and~\ref{fig:DC_0km} present the optimal pump power and extraneous count sensitivity analysis for the case without terrestrial fibre, corresponding to a scenario where the user is co-located with the OGS. The absence of the 10~km fibre---and thus the loss and dispersion it brings---significantly increases the key length while also expanding the admissible regions of pump power and extraneous counts compared to Figures~\ref{fig:optim} and~\ref{fig:DC}. As a result, the minimum elevation angle required for key generation is reduced, leading to longer overpass durations for a given satellite pass, as well as an increased number of viable overpasses due to the wider ground coverage enabled by the lower elevation threshold. Furthermore, the maximum tolerable extraneous count rate before key generation ceases is significantly increased, allowing for a greater robustness to background light and detector noise, which is particularly advantageous for practical implementations where ambient conditions cannot be fully controlled. However, this improved performance comes at the cost of reduced flexibility in user location. Introducing fibre links, while degrading performance through added loss, enables greater terrestrial coverage and relaxes constraints on user location relative to the OGS. This highlights a trade-off between system performance and network accessibility: increasing fibre length enhances deployment flexibility but progressively tightens the operational margins for successful key generation.

%

%--

\subsection{Relating Elevation to Ground Track Offset}
\label{supp1}

\begin{figure}[H]
    \centering
    \includegraphics[width=0.8\linewidth]{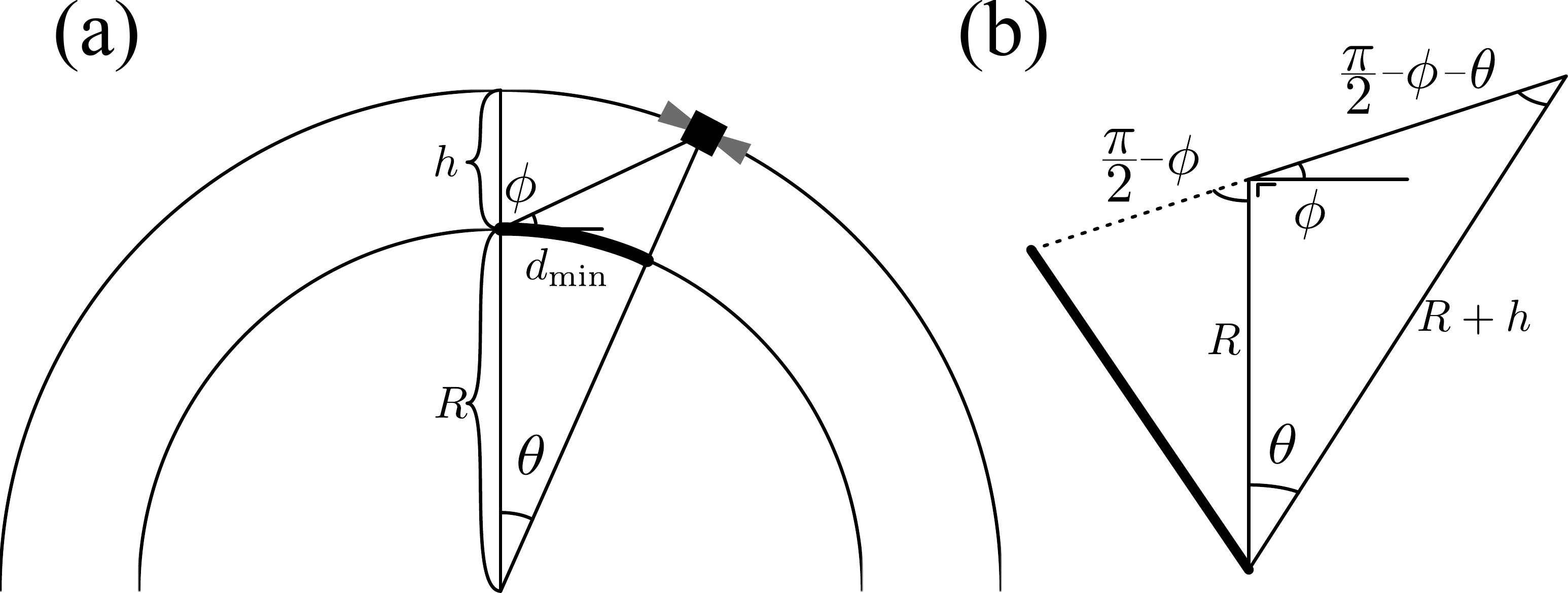}
    \caption{\textbf{General orbital geometry:} (a) A cross-section of the Earth, showing the satellite orbiting at an altitude $h$, and passing over the OGS with a ground-track offset $d_\text{min}$ at an angle $\phi$ with the horizon. 
    The ground track offset, illustrated by the bold segment, is determined from $R \theta$.
    (b) Illustration of how Eq.~\eqref{lawsin} may be obtained by using the law of sines in the triangle defined by the OGS, satellite and Earth centre.}
    \label{fig:sphere_angles}
\end{figure}

\noindent
We derive the relation between the ground-track $d_\text{min}=R\theta$ offset and elevation angle $\phi$ from the schematic illustrated in Figure~\ref{fig:sphere_angles}(a).
We focus on the triangle formed between the OGS, the satellite and the Earth's centre.
Two vertices are of known length, and two angles are the ones we are trying to relate. 
In order to be able to use the law of sines, we express the third angle as a function of the other two, and equate the projections of both known vertices under their respective angles, corresponding to the bold line in Figure~\ref{fig:sphere_angles}(b):
\begin{align}
    R\sin\left(\frac{\pi}{2}-\phi\right) = (R+h)\sin\left(\frac{\pi}{2}-\phi-\theta\right), 
\label{lawsin}
\end{align}
which may be reworked to Eq.~\ref{eq:earth}.

%

%--

\subsection{Cut-off Elevation angle }
\label{supp3}

\noindent
We outline a method to optimise the secret key length by adjusting the size of a single block corresponding to data collected during an overpass. The time-dependent errors are not uniformly distributed throughout the raw key obtained during a satellite pass. Previous work has shown that finite-key performance can be improved by excluding the highest-error portion of the data from the key generation process. This effect is particularly relevant for up-link architectures, which operate firmly in the finite-key regime. Due to the orbital geometry, the error rate decreases monotonically with increasing elevation angle. Optimising the block size therefore corresponds to introducing a cut-off elevation angle below which all detections are discarded.

\begin{figure}[h!]
    \centering    \includegraphics[width=0.7\linewidth]{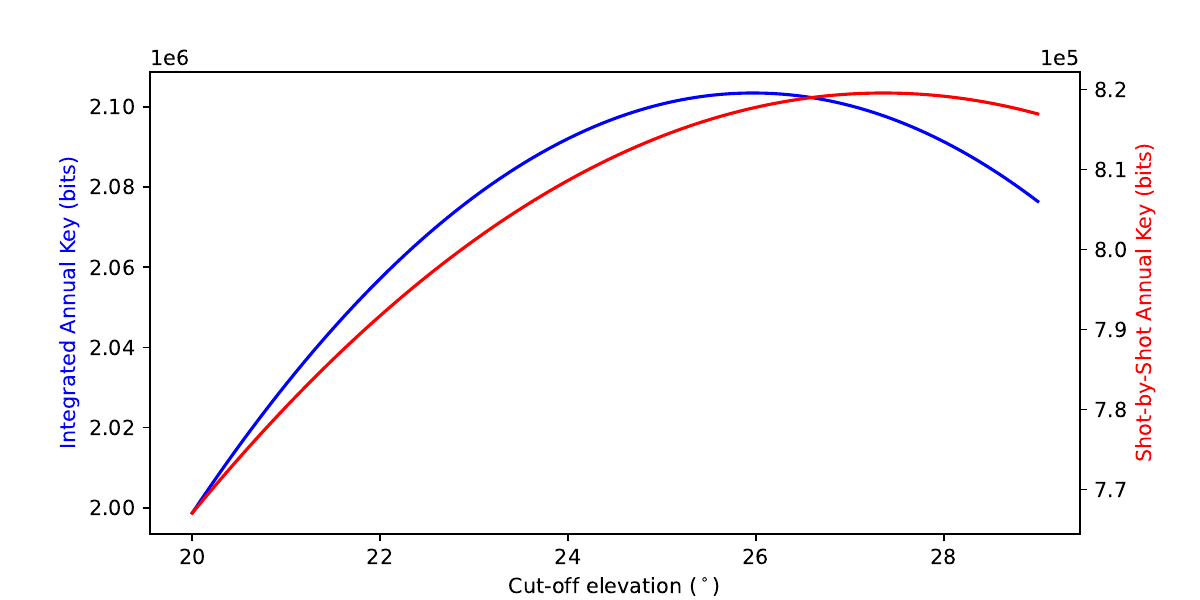}
    \caption{\textbf{Annual key length with cut-off elevation angle optimisation:} Annual key as a function of the cut-off elevation angle, below which data collected during a satellite overpass is excluded from key generation. For 10~km of fibre and a satellite at altitude $h=550$~km with an aperture of $R_\mathsf{X}=25$~cm, we compare the integrated annual key, where raw data are accumulated prior to key extraction, and the shot-by-shot annual key, where each overpass is processed independently. An optimal cut-off angle is observed in both cases, reflecting the trade-off between retaining additional signals and excluding high-error data acquired at low elevations. The optimum occurs at a higher elevation angle for the shot-by-shot annual key, as the smaller finite-key blocks are less tolerant to the increased QBER associated with low-elevation regions of the overpass.}
    \label{fig:cutoff}
\end{figure}

Figure~\ref{fig:cutoff} illustrates the integrated and shot-by-shot annual key rates for 10~km of fibre and a satellite at altitude $h=550$~km with an aperture of 25~cm as a function of this cut-off angle. In both cases, the annual key exhibits a broad optimum, indicating that the additional signals obtained at low elevations do not necessarily compensate for the accompanying increase in QBER. The optimal cut-off angle is also higher for the shot-by-shot annual key, since the smaller finite-key blocks are less tolerant to the increased proportion of errors present at low elevations. Consequently, finite-size effects favour a more selective use of overpass data, prioritising lower-error detections obtained near the peak of the pass.

\end{document}